\newif\ifPNAS\PNASfalse

\ifPNAS
  \documentclass[letterpaper]{article}
\else
  \documentclass[letterpaper, floatfix, twocolumn, showkeys]{revtex4}
\fi

\newcommand{\pnas}[1]{\ifPNAS#1\fi}
\newcommand{\orig}[1]{\ifPNAS\relax\else#1\fi}

\usepackage{graphicx}
\usepackage{psfrag}
\orig{\usepackage{natbib}}
\usepackage{color}
\usepackage{amsmath}
\usepackage{url}
\pnas{\usepackage{txfonts}}
\pnas{\usepackage[T1]{fontenc}}
\pnas{\usepackage[nolists,nomarkers]{endfloat}}
\pnas{}

\pnas{\newcommand{\citet}{\cite}\newcommand{\citep}{\cite}}

\pnas{\linespread{1.6}}

\begin{document}

\title{Modeling innovation by a kinetic description 
  of the patent citation system}
\date{\today}
\orig{\author{G\'abor Cs\'ardi}
\email[Corresponding author, ]{csardi@rmki.kfki.hu}
\affiliation{Department of Biophysics, KFKI Research Institute for
  Particle and Nuclear Physics of the Hungarian Academy of Sciences,
  Budapest, Hungary}
\affiliation{Center for Complex Systems Studies, Kalamazoo College,
  Kalamazoo, MI 490006, USA}
\author{Katherine J. Strandburg}
\affiliation{DePaul University -- College of Law, Chicago, IL 60604,
  USA}
\author{L\'aszl\'o Zal\'anyi}
\affiliation{Department of Biophysics, KFKI Research Institute for
  Particle and Nuclear Physics of the Hungarian Academy of Sciences,
  Budapest, Hungary }
\affiliation{Center for Complex Systems Studies, Kalamazoo College,
  Kalamazoo, MI 490006, USA}
\author{Jan Tobochnik}
\affiliation{Department of Physics, Kalamazoo College, Kalamazoo, MI
  490006, USA} 
\affiliation{Center for Complex Systems Studies, Kalamazoo College,
  Kalamazoo, MI 490006, USA}
\author{P\'eter \'Erdi}
\affiliation{Department of Biophysics, KFKI Research Institute for
  Particle and Nuclear Physics of the Hungarian Academy of Sciences,
  Budapest, Hungary }
\affiliation{Center for Complex Systems Studies, Kalamazoo College,
  Kalamazoo, MI 490006, USA }
}

\pnas{\author{G\'abor Cs\'ardi$^{*\dagger\P}$ \and 
  Katherine Strandburg$^\ddagger$ \and
  L\'aszl\'o Zal\'anyi$^{*\dagger}$ \and 
  Jan Tobochnik$^{\S\dagger}$ \and 
  P\'eter \'Erdi$^{*\dagger}$}}

\pnas{\maketitle}

\pnas{
\begin{center}
\vspace*{-.5cm}
$^*$ Department of Biophysics, KFKI Research Institute for
  Particle and Nuclear Physics of the Hungarian Academy of Sciences,
  29-33 Konkoly-Thege Road, Budapest, H-1121, Hungary\\
$^\dagger$ Center for Complex Systems Studies, Kalamazoo College,
  1200 Academy Street, Kalamazoo, MI 490006, USA\\
$^\ddagger$ DePaul University -- College of Law, 25 East Jackson
  Boulevard, Chicago, IL 60604,  USA\\
$^\S$ Department of Physics, Kalamazoo College, Kalamazoo, MI
  490006, USA \\
$^\P$ Corresponding author, phone: +36 (1) 392-2222, 
  fax: +36 (1) 392-2742, email: csardi@rmki.kfki.hu
\end{center}
}

\pnas{\noindent Classification: social sciences, economic sciences}

\pnas{\noindent Number of pages: 18\\
      Number of figures: 5\\
      Number of tables: none}

\pnas{\noindent Number of words in the abstract: 178\\
  \enlargethispage{1cm}%
  Total number of characters: 43775}

\pnas{\noindent Contributions of the authors: 
  G.Cs., K.S., L.Z., J.T. and P.\'E. performed research, 
  contributed new analytic tools and wrote the paper.}

\pnas{\newpage}

\begin{abstract}
This paper reports results of a network theory approach to the study
of the United States patent system.  We model the patent citation
network as a discrete time, discrete space stochastic dynamic system.
From data on more than 2 million patents and their citations, we
extract an attractiveness function, $A(k,l)$, which determines the
likelihood that a patent will be cited.  $A(k,l)$ is approximately
separable into a product of a function $A_k(k)$ and a function $A_l(l)$,
where $k$ is the number of citations already received (in-degree) and $l$
is the age measured in patent number units.  $A_l(l)$ displays a peak at
low $l$ and a long power law tail, suggesting that some patented
technologies have very long-term effects.  $A_k(k)$ exhibits super-linear
preferential attachment. The preferential attachment exponent has been
increasing since 1991, suggesting that patent citations are
increasingly concentrated on a relatively small number of patents.
The overall average probability that a new patent will be cited by a
given patent has increased slightly during the same period.  We
discuss some possible implications of our results for patent policy.
\end{abstract}

\pnas{\newpage}
\orig{\maketitle}

\section{Introduction}%
\label{sec:intro}

Innovation plays a key role in economic development and the patent
system is intended (and Constitutionally required in the United
States) to promote innovative progress.  The patent system promotes
innovation by giving inventors the power to exclude others from using
their inventions during the patent term.  The power to exclude is a
double-edged sword, however, benefiting the original inventor, but
imposing costs on later innovators seeking to build on past
inventions.  Thus, the proper design of the patent system is an
important matter -- and a matter of considerable current debate.  See,
e.g., \cite{jaffe04,ftc03,merrill04}. Advances in computer technology and the
availability of large patent databases have recently made it possible
to study aspects of the patent system quantitatively.  To date the
empirical analysis of the patent system has been undertaken by
economists and some legal scholars.  See, e.g.,
\cite{jaffe02,allison04,moore05}.  Because 
patents and the citations between them can be conceptualized as a
growing network, however, techniques from statistical physics that
have been used in the study of complex networks can be usefully
applied to the patent citation network \cite{albert02,newman03}.  
In this paper we
present what we believe to be the first results of a network theory
approach to the patent system.  We explore the kinetics of patent
citation network growth and discuss some possible implications for
understanding the patent system. 

The paper is organized as follows:  In Section II we provide
background on the United States patent system and describe the
citation data that is used in this study.  In Section III we describe
a general framework for modeling the kinetics of citation networks,
define an ``attractiveness function'' for the evolving
network and introduce an iterative method for extracting the
attractiveness function from the data.  In Section IV we apply this
approach to analyze the US patent citation network and explore the
changes in the kinetics from 1976 to 2000. In Section V we discuss
some possible implications of our results, and mention directions for
future research.  
\section{Patentological background}%
\label{sec:background}

While a similar approach could be applied to many patent systems,
including the very important European and Japanese patent systems, we
begin our analysis with the United States patent system for which an
extensive database of citations has been made available through the
work of economists Hall, Jaffe, and Trajtenberg \cite{hall03}.  

An application for a U.S. Patent is filed in the U.S. Patent and
Trademark Office (USPTO).  A patent examiner at the USPTO determines
whether to grant a patent based on a number of criteria, most of
important of which for present purposes are the requirements of
novelty and non-obviousness with respect to existing technology.  Once
a patent is issued by the USPTO, it is assigned a unique patent
identification number.  These numbers are sequential in order of
patent grant.   

Novelty and nonobviousness are evaluated by comparing the claimed
invention to statutorily defined categories of ``prior
art'', consisting in most cases primarily of prior patents.
Patents are legally effective only for a limited term (currently
twenty years from the date of application), but remain effective as
``prior art'' indefinitely.  Inventors are required to
provide citations to known references that are
``material'' to patentability, but are not required to
search for relevant references (though they or their patent attorneys
often do so).  During consideration of the application, patent
examiners search for additional relevant references.   

Patent citations reflect potential prior art that was considered by
the examiner.  They thus reflect the judgment of patentees, their
attorneys, and the USPTO patent examiners as to the prior patents that
are most closely related to the invention claimed in an application.
Patent citations thus provide, to some approximation, a
``map'' of the technical relationships between patents
in the U.S. patent system.  This ``map'' can be
represented by a directed network, the nodes being the patents and the
directed edges the citations.  Our research uses a statistical physics
approach inspired by studies of other complex networks to attempt to
gain insight from that ``map''. 

The patent database we use for the analysis in this paper was created
by Hall, Jaffe and Trajtenberg based on data available from the US
Patent Office \cite{hall03}. It is available online at
\url{http://www.nber.org/patents/}.  The database contains data from over 6
million patents granted between July 13, 1836 and December 31, 1999
but only reflects the citations made by patents after January 1, 1975:
more than 2 million patents and over 16 million citations.  Citations
made by earlier patents are also available from the Patent Office, but
not in an electronic format.  The Hall, Jaffe and Trajtenberg database
also contains additional data about the included patents, which is
described in detail in \cite{hall03}. 

\section{Modeling patent citation networks}%
\label{sec:modeling}


\subsection{Defining the model framework}

In this section we define the mathematical model framework we will use
for studying patent citations. This framework is a discrete time,
discrete space stochastic dynamic system.  Time is measured in patent
number units.  We often ``bin'' the data from a
range of patent numbers to obtain sufficient statistics for the
analysis. In our model, each patent is described by two variables: 

\begin{enumerate}
\item $k$, the number of citations it has received up to the current time
  step and 
\item $l$, the age of the patent, which is simply the difference
  between the current time step (as measured in patent numbers) and
  the patent number.  Because a given patent may cite more than one
  other patent, several citations may be made in one time step.
\end{enumerate}

These two variables define what we call the
``attractiveness'' of a patent, $A(k,l)$ which determines
the likelihood that the patent will be cited when the next citation is
made.  In every time step the probability that an older patent will be
cited is proportional to the older patent's attractiveness
multiplied by the number of citations made in that time step.  We find
that this simple model gives a very good approximation of the observed
kinetics of the growth of the patent citation network. 

More formally, the state of the system is described by $k_i(t)$ and
$l_i(t)$, $(1<i<N)$, where $N$ is the patent number of the last
patent studied and $k_i(t)$ and $l_i(t)$ are the in-degree and
age, respectively, of patent $i$ at the beginning of time step $t$. The
attractiveness of any node with in-degree $k$ and age $l$ is denoted by
$A(k,l)$.  $A(k,l)$ is defined such that the probability that node $i$ will
be cited by a given citation $e$ in time step $t$ is given by 

\begin{equation}
P[e \text{ cites node } i ]=\frac{A(k_i(t), l_i(t))}{S(t)},
\end{equation}
where $S(t)$ is the total attractiveness of the system at time step $t$.
\begin{equation}
S(t)=\sum_{j=1}^{t} A(k_j(t), l_j(t)).  
\end{equation}

The total probability that node $i$ will be cited in time step $t$ is thus
$E(t) A(k_i(t),l_i(t))/S(t)$, where $E(t)$ is the number of citations made
by patent $t$.  $A(k,l)$ and $S(t)$ are defined up to an arbitrary
normalization parameter.  To normalize, we arbitrarily define
$A(0,1)=1$.  With this normalization, $S(t)$ is the inverse probability
that a ``new'' node, with $k=0$ and $l=1$, will be cited by a
given citation during the next time step.   

The $A(k,l)$ function determines the evolution of the network.  It
describes the average citation preferences of the citing patents (the
inventors and patent examiners in reality).  In this study, we measure
and analyze $A(k,l)$ for the United States patent system during the
time period covered by our data.  We find first that the parameterization by
$k$ and $l$ consistently describes the average kinetics of the patent
citation network.  Of course, underlying patent citations are patentee
and patent examiner evaluations of the significance of the cited
patent and the technological relationship between the citing and cited
patents that our probabilistic approach cannot capture.  The way in
which these ``microscopic dynamics'' are translated into
the average behavior that we observe remains an open question.   

In the following part of this section we will explain our method for
measuring the $A(k,l)$ and $S(t)$ functions for a given network.  We
believe that this method may be usefully applied to other networks as
long as the necessary data is available. 

\subsection{Measuring the attractiveness function}

Let us assume that edges are added to the system one after another
in a fixed order; if two edges are added in the same time step (i.e.,
by the same citing patent), their order is fixed arbitrarily for the
measurement. Let $e$ be an edge and let $c_e(k,l)$ be indicator random
variables, one for each $(e,k,l)$ triple, 
$(1 < e < E_\text{tot}, k \ge 0, l > 0)$, 
where $E_\text{tot}$ is the total number of edges in the
system. $c_e(k,l)$ is one 
if and only if edge $e$ cites a $(k,l$) node (i.e., a node having
in-degree $k$ and age $l$) and zero otherwise. The probability that
edge $e$ cites a $(k,l)$ node, i.e., that $c_e(k,l)$ is one, is thus
given by 
\begin{equation}
P[c_e(k,l)=1]=\frac{N(t(e),k,l)A(k,l)}{S(t(e))}
\end{equation}
where $t(e)$ is the time step during which edge $e$ is added,
$S(t(e))$ is the total attractiveness of the system right before
adding edge $e$, 
and $N(t(e),k, l)$ is the number of $(k,l)$ nodes in the network right
before adding edge $e$. We thus have a formula for $A(k,l)$: 
\begin{equation}
A(k,l)=\frac{P[c_e(k,l)=1]S(t(e))}{N(t(e),k,l)}
\label{eq:akl}
\end{equation}

In (\ref{eq:akl}) it is easy to determine $N((t(e),k,l)$ for any 
$(e,k,l)$, but $S(t(e))$ is unknown.  Moreover, we have only a single
experiment for $c_e(k, l)$ which is not enough to approximate 
$P[c_e(k,l)=1]$ properly. 
To proceed further, let us define a new set of random variables, each
of which is a simple transformation of the corresponding $c_e(k, l)$
variable: 
\begin{equation}
  A_e(k,l)=\frac{c_e(k,l) S(t(e))}{N(t(e),k,l)}, \quad \text{if
    $N(t(e),k,l)>0$} 
\end{equation}
If $N(t(e),k,l)=0$ then $A_e(k,l)$ is not defined. It is easy to see
that the expected value of any $A_e(k,l)$ variable (if defined) is 
$A(k,l)$; thus we can approximate $A(k,l)$ by
\begin{equation}
  \bar{A}(k,l)=\frac{1}{E(k,l)} \sum_{e=1}^{|E_\text{tot}|}\frac{\bar{c}_e(k,l)
    S(t(e))}{N(t(e),k,l)}
  \label{eq:abar}
\end{equation}
Here $E(k,l)$ is the number of edges for which $N((t(e),k,l))>0$ for
any $t(e)$, and $\bar{c}_e(k, l)$ is the realization of $c_e(k,l)$ in
the network being studied.

To calculate this approximation for $A(k,l)$ we need to determine
$S(t(e))$, which itself is defined in terms of $A(k,l)$.  To determine
$A(k,l)$ and $S(t(e))$ self-consistently, we use the following iterative
approach: 

\begin{enumerate}
\item First we assume that $S_0(t)$ is constant, and use (\ref{eq:abar})
  to compute $A_0(k, l)$, normalizing the values such that $A_0(0,1)=1$.
\item Then we calculate $S_1(t)$ for each $t$ based on $A_0(k,l)$ and
  use this to determine $A_1(k,l)$.
\item We repeat this procedure until the difference between $S_n(t)$
  and $S_{n+1}(t)$ is smaller than a given small $\epsilon$ for all
  $t$.
\end{enumerate}

To check this iterative method, we have applied it to various
well-known models of growing networks, such as the Barabasi-Albert
model \cite{barabasi99}.  In these tests the method yielded the
correct form of the 
$A(k,l)$ function, which, for the BA-model, for example, is 
$A(k,l) = k+a$.  While these tests gave very good agreement overall
they also suggested that the method cannot accurately measure the
attractiveness of young nodes (small $l$) with high in-degree (high
$k$), as these occur very rarely in any finite sample network.


\section{Results}%
\label{sec:results}

\subsection{The attractiveness function}


\begin{figure}
\centering
\orig{%
  \begin{psfrags}
  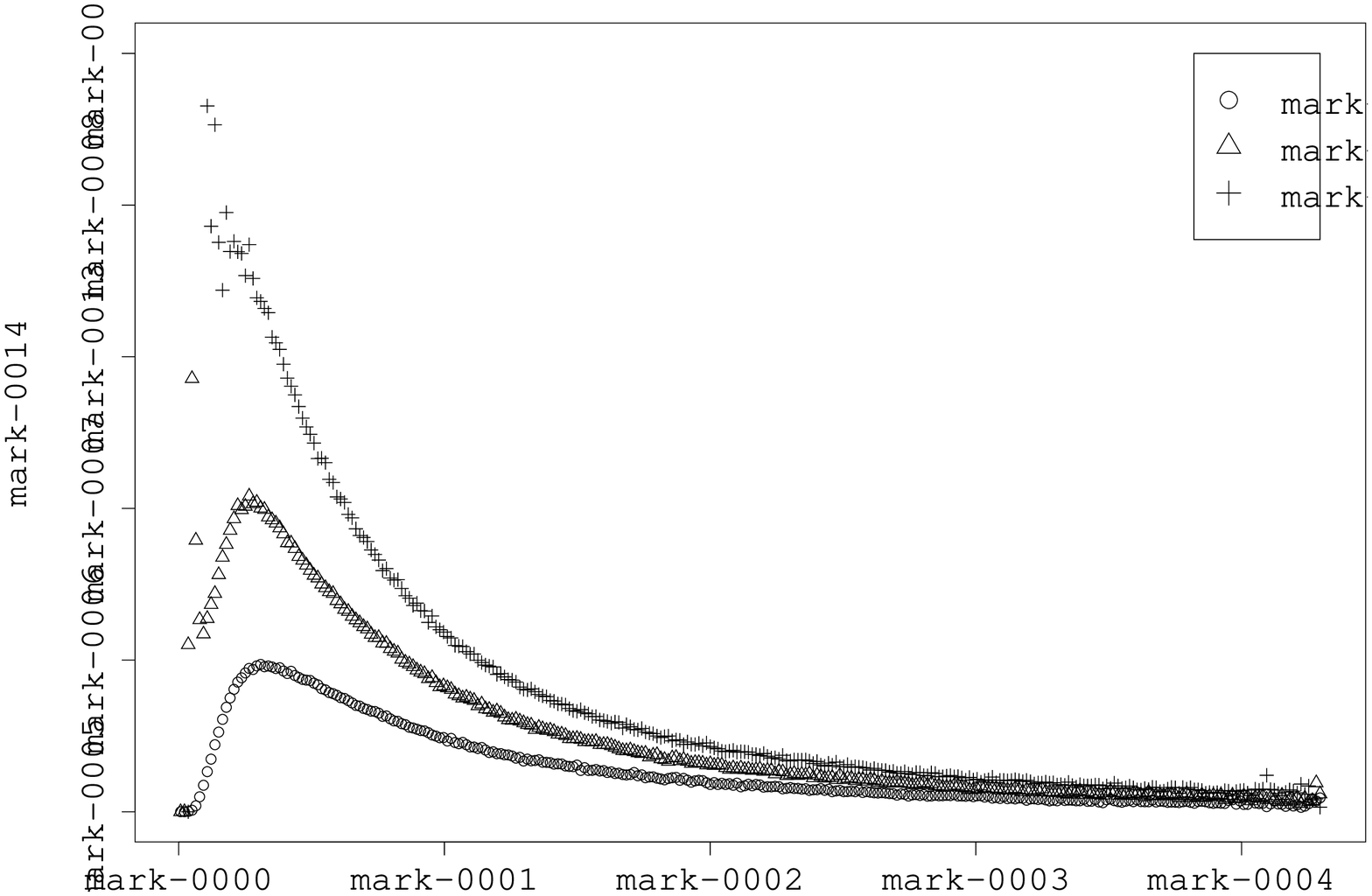
  \includegraphics[width=8.7cm]{042705-6.eps}
  \end{psfrags}
\\[12pt]}
\orig{%
  \begin{psfrags}
  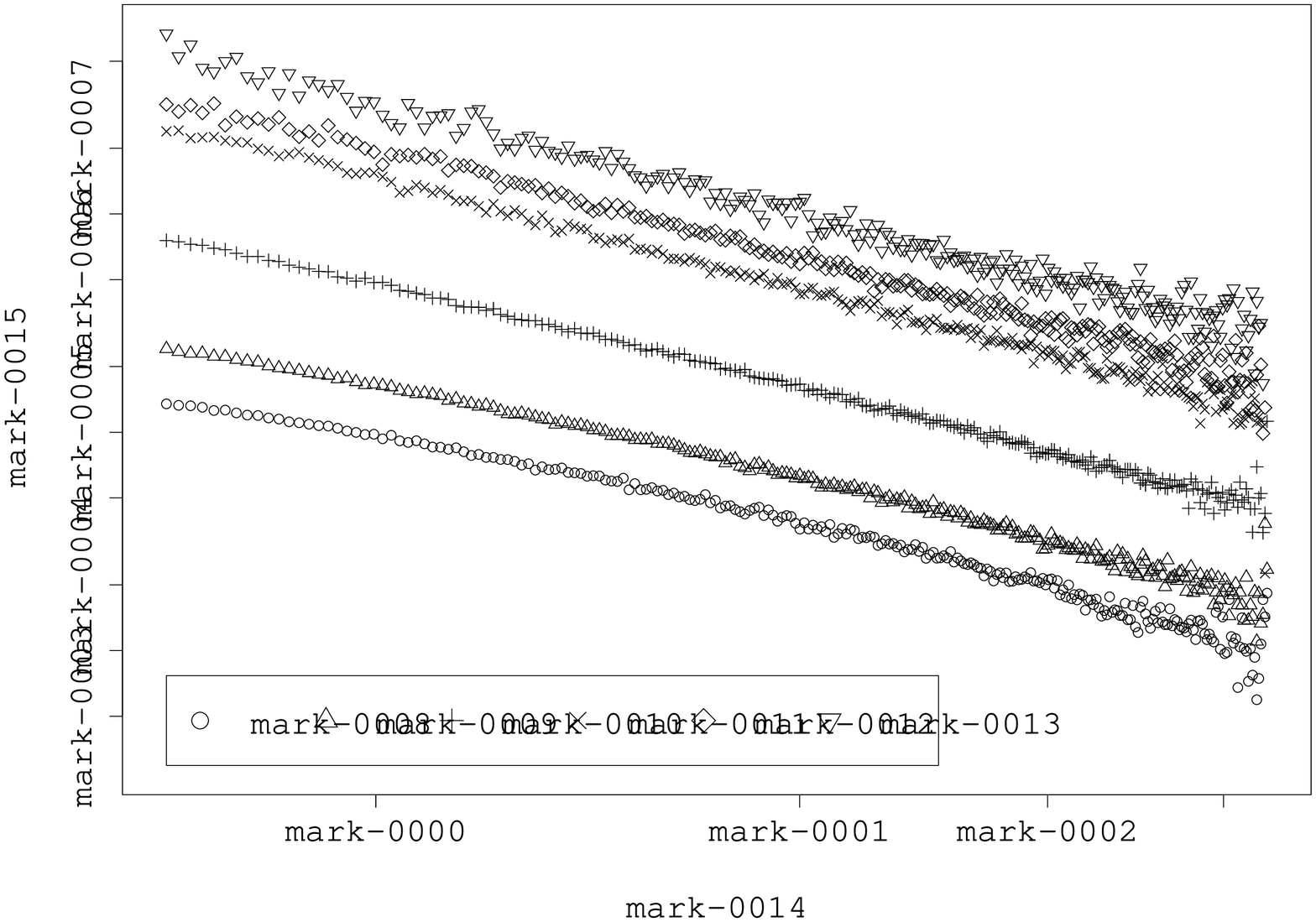
  \includegraphics[width=8.7cm]{042705-8.eps}
  \end{psfrags}
\\}
\caption{The measured attractiveness $A(k,l)$ as a function of age $l$
  for various fixed values of in-degree, $k$.  The bottom figure shows
  only the decreasing tail on log-log scales. 
  }
\label{fig:aging}
\end{figure}

\begin{figure}
\centering
\orig{%
  \begin{psfrags}
  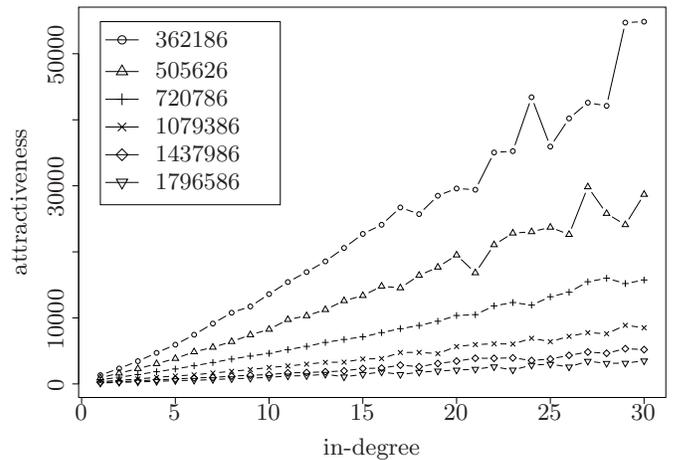
  \includegraphics[width=8.7cm]{042705-7.eps}
  \end{psfrags}
\\}
\caption{The measured attractiveness $A(k,l)$ as a function of
  in-degree, $k$,  for various fixed values of age, $l$. }
\label{fig:prefatt}
\end{figure}

The analysis method described in the previous section was applied to
the patent citation network and the forms of $S(t)$ and $A(k,l)$ were
determined. Figures \ref{fig:aging} and \ref{fig:prefatt} show
sections of the $A(k,l)$ function and 
Figure \ref{fig:S(t)} shows $S(t)$ (which is also the inverse of the
probability 
that a ``new'' node with $k=0$ and $l=1$ will be cited).
For all the figures in this paper we have binned the age values into
300 bins, each containing 7172 patents.  Ages and times are measured in
patent number units.  Figures \ref{fig:aging} and \ref{fig:prefatt}
suggest that, for the patent 
network, the effects of in-degree and age can be separated to good
approximation and that $A(k,l)$ can be written approximately in the
form 
\begin{equation}
\label{eq:separate}
A(k,l) = A_k(k)\cdot A_l(l).
\end{equation}
While this is a reasonable and useful approximation, it is also clear
that it is only approximately true. e.g., $A(0, \cdot)$ decays faster
than $A(30, \cdot)$, see the second plot in Figure \ref{fig:aging}.


The measured $A_l(l)$ function for the patent citation network has two
major features -- a peak at approximately 200,000 patent numbers and a
slowly decaying tail.  (The very large absolute values of $A_l(l)$ are
a result of the normalization, $A(0,1)=1$, and are of no independent
significance.)  The peak at 200,000 patent numbers corresponds to a
large number of what might be called ``ordinary'',
relatively short-term citations.  In 1998--1999, 200,000 patent numbers
corresponded to about 15 months.  The tail is best described by a
power-law decay: $A_l(l) \sim l^{-\beta}$ with $\beta\approx 1.6$.
The observation 
of this power law decay is an important result.  It indicates that
while typical citations are relatively short-term, there are a
significant number of citations that occur after very long delays.
Very old patents are cited, suggesting that the temporal reach of some
innovations, which perhaps can be described roughly as
``pioneer'', is very long indeed.  Moreover,
because $A_l(l)$ is approximately independent of $k$ -- i.e.,
approximately the same power law decay is observed even for small $k$ --
the power law 
tail of $A_l(l)$ demonstrates that there is a significant possibility
that patents that have gone virtually un-cited for long periods of
time will reemerge to garner citations.  This slow power law decay of
$A_l(l)$ thus suggests the unpredictability of innovative progress. 

The measured $A_k(k)$ function increases monotonically with $k$, as
Figure \ref{fig:prefatt} suggests.  Higher in-degree always means
higher attractiveness. 
Since the citation probability is proportional to the attractiveness,
this means that the well-known preferential attachment, or
``rich get richer'' effect is at work here -- the more
citations a patent has received, the more likely it is to receive
another.  The functional form of $A_k(k)$ is a power law over the entire
range of $k$ values.  $A_k(k)\sim k^\alpha+a$, where
$\alpha=1.19$  and $a=1.11$.  We estimated
these parameters using the smaller values of $k$, for which we have more
data, and the first 100 age bins.  We then checked the results by
comparing with more extensive fits.  

Preferential attachment and its variations are well studied, see the
reviews by Albert and Barabasi \cite{albert02} and by Newman
\cite{newman03}.  Linear preferential attachment ($\alpha=1$) without
aging has been shown to result in a degree distribution (frequency of
nodes with degree $k$) with a power law tail \cite{albert02,newman03}.
Krapivsky et al. \cite{krapivsky00} have studied nonlinear 
preferential attachment. In the model they studied there was no aging,
$A(k,l)=A_k(k)=k^\alpha+a$. For $a>1$, as is observed in the patent
citation network, their calculations predict a condensation of node
connectivity, in the sense that with high probability most of the
edges are connected to only a small number of nodes.  More
specifically, in their model, if $(m+1)/m < \alpha < m/(m-1)$ 
the number of nodes
with more than $m$ incoming edges is finite, even in an infinite
network. For the patent network $7/6 < \alpha < 6/5$ suggesting that, if
there were no aging, the number of patents receiving more than 6
citations would be very small, though those patents would account for
a large fraction of all of the citations.  Aging complicates this
picture, of course, and likely precludes a complete condensation onto
a few nodes.  However, the fact that the observed preferential
attachment is super-linear does indicate a tendency toward what might
loosely be called ``stratification''  -- many nodes with
very few citations and a few nodes with many citations. 

\subsection{The total attractiveness}

\begin{figure}
\centering
\orig{%
  \begin{psfrags}
  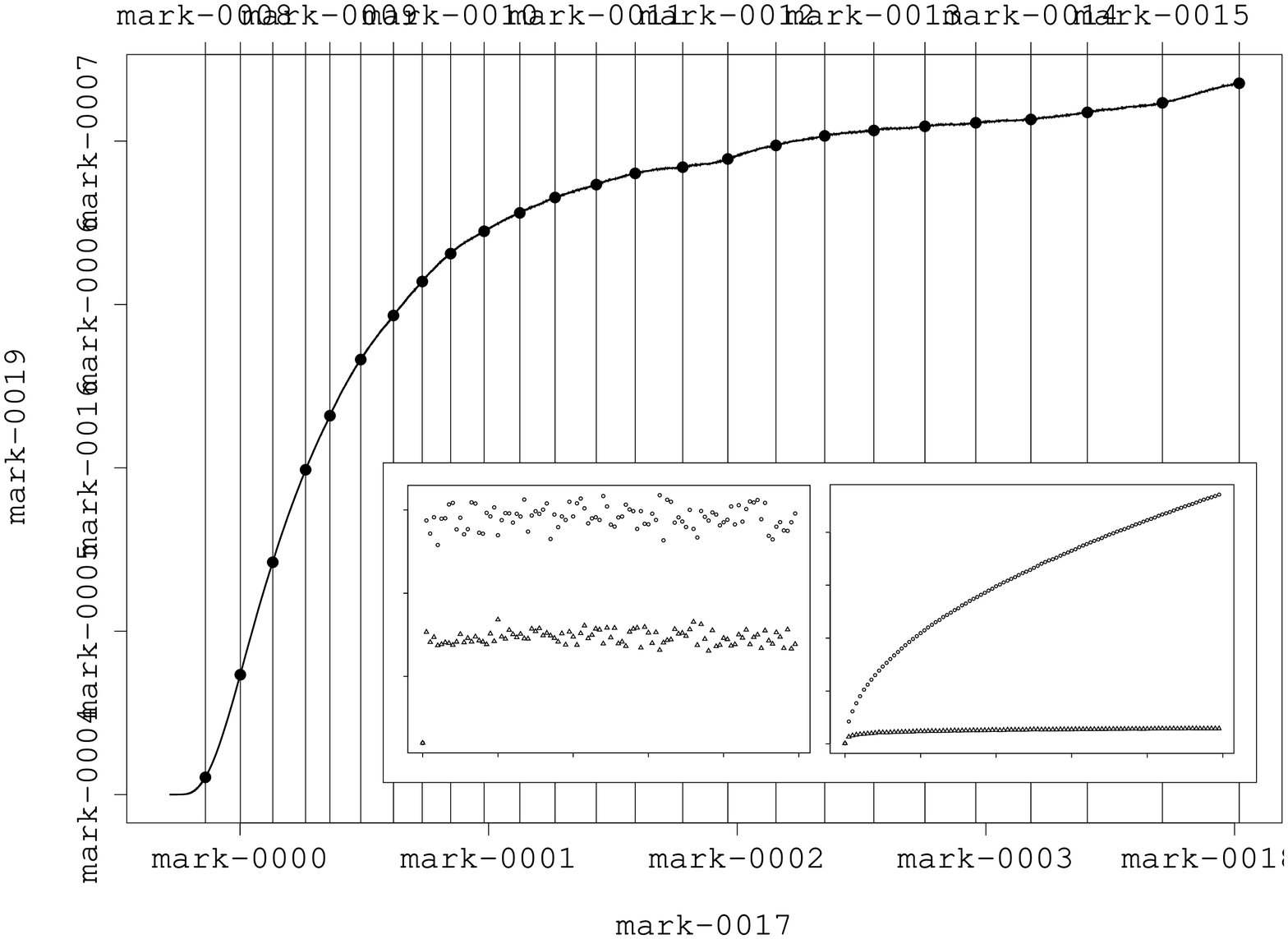
  \includegraphics[width=8.7cm]{042705-4.eps}
  \end{psfrags}
}
\caption{The total attractiveness $S(t)$ of the patent network versus
  time in units of patent numbers. For ease of reference the time in
  years is indicated by filled circles and vertical lines. The left
  and right figures in the inset show the total attractiveness of
  Model 1 and Model 2, as discussed in the text surrounding equation
  (8) and (9).  The left plot shows results of simulations for 
  $\lambda=1/50$ (circles) and 1/5 (triangles).  The right plot shows results
  of simulations for $\mu=1/2$ (circles) and 1 (triangles). The
  simulations were run for 10,000 nodes.
}
\label{fig:S(t)}
\end{figure}

The total attractiveness function, $S(t)$, (see Fig. \ref{fig:S(t)}) of the US
patent system increases with time.  The initial steep increase is only
a finite size effect and comes from the fact that the citations made
by pre-1975 patents are missing from our database. From about 1984 on,
however, $S(t)$ displays a slow but steady increase.  One way to
interpret this increase is that the probability that a patent will be
cited by a given citation (which is proportional to $1/S(t)$) is
decreasing as the size of the network increases.  The decrease is
determined in part by the rate at which patents age, which determines
the number of patents ``available'' for citation.   

To better understand the behavior of the $S(t)$ function, we simulated
two simple growing network models with two different
``toy'' $A(k,l)$ functions, with linear preferential
attachment and two different forms of age dependence: 
\begin{align}
\text{ Model 1 }\quad A_1(k,l)& = (k+1) \cdot e^{-\lambda l} \text{ and}
\label{eq:model1}\\
\text{ Model 2 }\quad A_2(k,l)& = (k+1) \cdot l^{-\mu}.
\label{eq:model2}
\end{align}

$S_1(t)$ and $S_2(t)$ were determined for these models, see
Fig. \ref{fig:S(t)}, inset. In these models a single edge was added to
the network at each time step. 

When the attractiveness function decays exponentially with age (Model
1), the total attractiveness fluctuates around a constant value,
which is determined by $\lambda$ and is independent of the system size.
In Model 1 the probability that a new node will be cited by a
particular citation is thus always the same.  The exponentially
decaying age dependence means that the effect of very old nodes is
negligible; there is effectively a constant-sized
``band'' of recent nodes that remain
``citable''.  In a patent citation system, such an
exponential age dependence would imply, contrary to our observations,
that the importance of innovations is short-lived.  

In Model 2, the behavior of $S_2(t)$ depends on $\mu$.  If $\mu$ is
below a limit value $\mu_1$ (which is about 1), $S_2(t)$ is sharply
increasing.  If $\mu$ is between $\mu_1$ and another limit, $\mu_2$
(which is about 1.5) $S_2(t)$ increases slowly.  When $\mu$ is higher
than $\mu_2$, $S_2(t)$ fluctuates around a constant value, as does
$S_1(t)$. 
Thus Model 2 exhibits a crossover from a regime of slowly decaying age
dependence in which old nodes remain influential, to a regime of more
rapidly decaying age dependence in which old nodes are
``forgotten''.  These results are in good agreement with
other theoretical studies about aging and preferential attachment; see
the work by Dorogovtsev and Mendes \cite{dorogovtsev00}, Zhu et
al. \cite{zhu03}, Klemm and Eguluz \cite{klemm02}. 

These simple models would suggest that the patent citation network
(which has an aging exponent of about 1.6 -- above $\mu_2$ -- should
have $S(t)$ roughly constant in time.  Indeed, a third toy model with
superlinear preferential attachment (exponent 1.2) and power-law aging
(exponent 1.6) displayed a roughly constant $S(t)$.  However, the
observed $S(t)$ for the patent system increases with time.

\begin{figure}
\orig{%
  \begin{psfrags}
  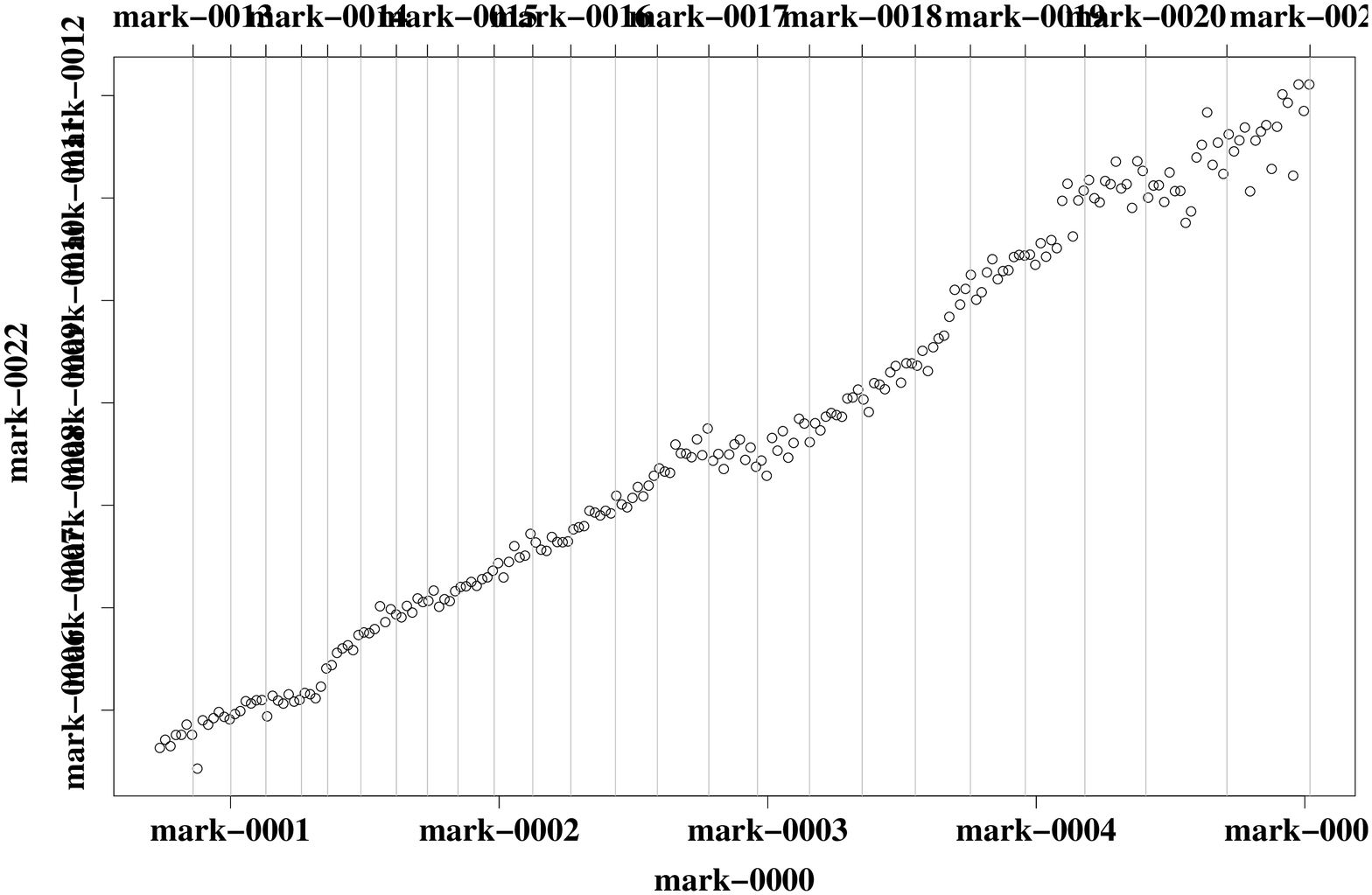
  \includegraphics[width=8.7cm]{072705-5.eps}
  \end{psfrags}
\\[12pt]}
\orig{%
  \begin{psfrags}
  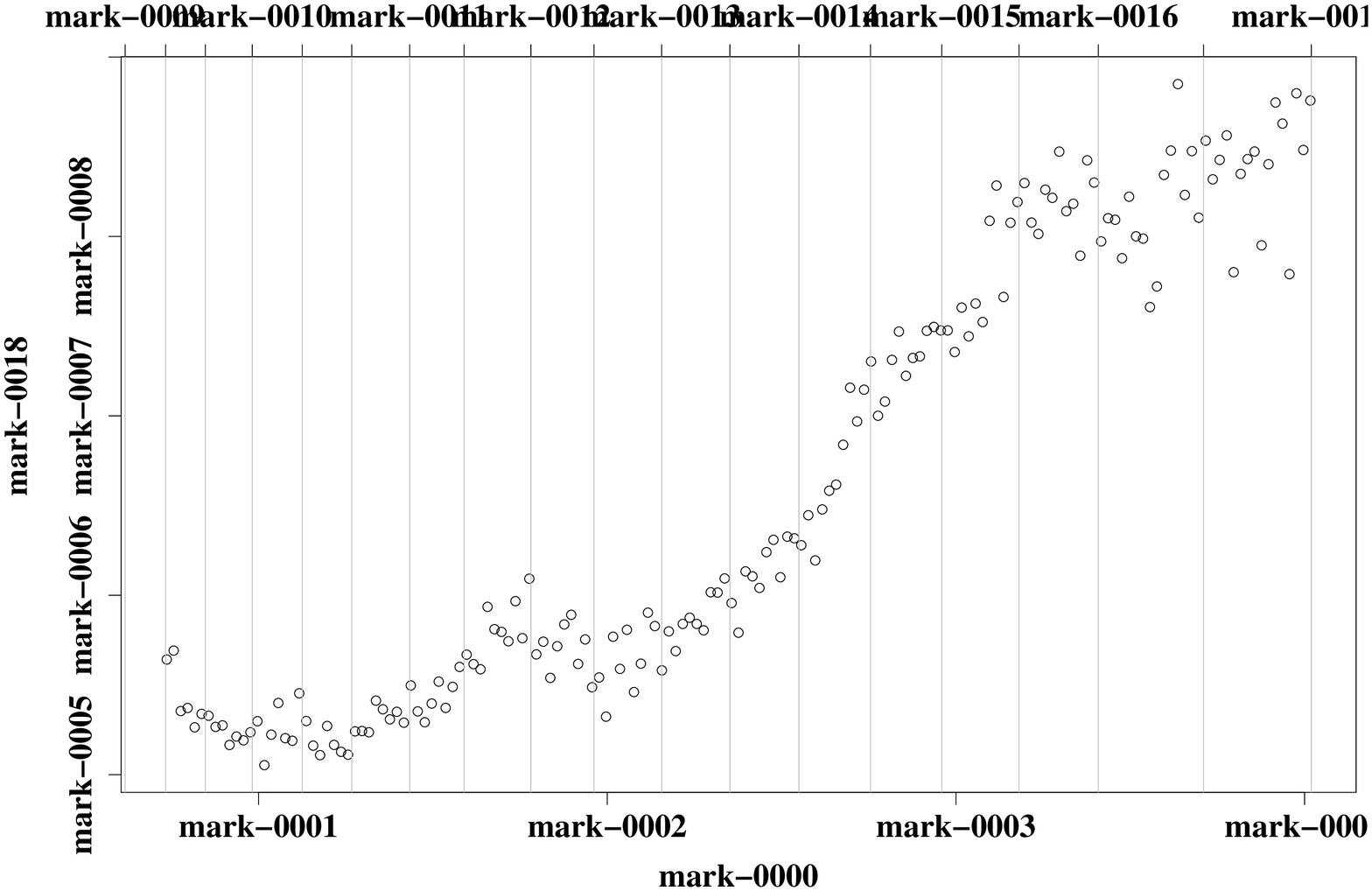
  \includegraphics[width=8.7cm]{072705-6.eps}
  \end{psfrags}
}
\caption{The top figure shows the number of citations made per patent,
  $E(t)$, as a function of time in units of patent number.  The bottom
  figure shows $E(t)/S(t)$, corresponding also to the probability that a
  new patent with $k=0$, $l=1$ will be cited, as a function of time in
  units of patent number.}
\label{fig:totalchange}
\end{figure}

The explanation for this apparent contradiction lies in the fact that
the average number of citations made by each patent (and hence, since
we measure time in units of patents, the number of citations made in
each time step, $E(t)$), has increased approximately linearly with time
in the real patent citation network, e.g., it was 4.69 in 1975 and
10.66 in 1999.  See Fig. \ref{fig:totalchange}. A modified simulation
with superlinear 
preferential attachment (exponent 1.2) and power-law aging (exponent
1.6), but with linearly increasing number of citations per time step
confirmed that, as in the real patent data, the increasing number of
citations made in each time step results in an increasing overall
attractiveness, $S(t)$. 

The probability that patent $i$ will be cited in a given time step (in
other words, by a particular patent rather than by a particular
citation) is  

\begin{equation}
P[k_i(t + 1)=k_i(t)+1]=E(t)\frac{A(k_i(t),l_i(t))}{S(t)}
\end{equation}

The probability that a new patent $(k=0,l=1)$ will be cited by the next
patent is thus given by $E(t)/S(t)$, which is shown in
Fig. \ref{fig:totalchange}.  From
this plot one can see that the increase in the number of citations
being made outweighs the increase in $S(t)$, so that the probability
that a new patent will be cited has increased over time, despite the
increasing $S(t)$.  Patents do not get ``lost in the
crowd'' the way we might have predicted from the simple models.
Instead, patentees and patent examiners have on average increased the
number of citations made by each patent to more than compensate for
the increasing $S(t)$. 

\subsection{Change in the patent system dynamics}


While it is well known that there has been a significant increase in
the number of US patents granted per year since 1984
\cite{jaffe04,hall05}, the 
underlying reason for this increase is not clear.  Has there simply
been an acceleration of technological development in the last twenty
years or has there been a more fundamental change in the patent
system, perhaps, as many have suggested, as a result of increased
leniency in the legal standard for obtaining a patent \cite{jaffe04}. A
complete answer to this question is far beyond the scope of the
present investigation.  However, our kinetic model does permit us to
ask whether there has been any deep change in the growth kinetics of
the patent citation network.  Because we measure time in units of
patent number, a mere acceleration of technological progress should
leave $A(k,l)$ unchanged in patent number ``time''.  A
change in $A(k,l)$ indicates some other source of change. 

\begin{figure}
\centering
\orig{%
  \begin{psfrags}
  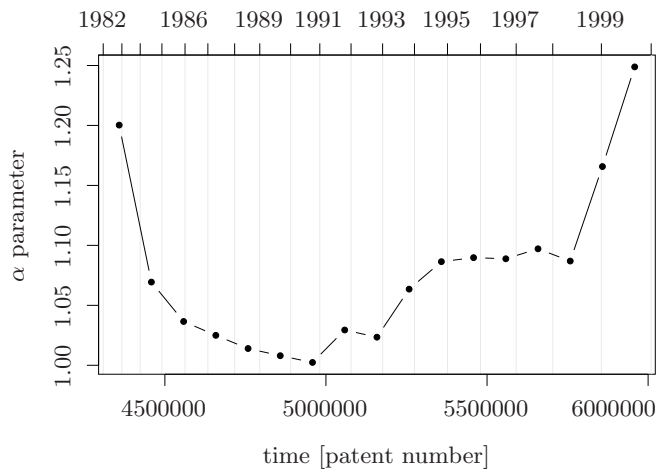
  \includegraphics[width=8.7cm]{050205-4.eps}
  \end{psfrags}
\\}
\caption{The measured value of $\alpha$ as a function of time,
  measured as described in the text.  The time in years is indicated
  as in Figure \ref{fig:totalchange}. 
}
\label{fig:change}
\end{figure}

Thus far, we have assumed a time-independent $A(k,l)$, which is
reasonably consistent with our observations.  In this section, we
relax this assumption to ask the more subtle question of whether there
has been a change in patent system kinetics over and above the
acceleration that is already reflected in our choice of time units.
Specifically, we allow $\alpha$ and $\beta$ to vary with time and ask
whether there has been a significant change in these parameters
between 1980 and 2000.  

To answer this question we measured the parameters of the system as
functions of time. To perform the fits, we averaged over a
500,000-patent sliding time window and calculated the parameters after
every 100,000 patents.  The measured $\alpha$ parameters are plotted in
Figure \ref{fig:change}.  There is a significant variation over time.  The time
dependence of the important $\beta$ parameter was also explored, but no
significant time dependence was observed to within the statistical
errors. 

The plot for the $\alpha$ parameter shows that there are two regimes.
In the first regime, prior to about 1991, $\alpha$ is decreasing
with time, while in the second, starting around 1991, there is a
significant increase.   
As noted earlier, the $\alpha$ parameter has some very important consequences
for the growth of the network: the higher $\alpha$, the more
``condensed'' or ``stratified'' the
network will be. The increasing $\alpha$ in the patent citation network
indicates increasing stratification -- a smaller and smaller fraction
of the patents is receiving a larger and larger fraction of the
citations.  This change is not simply a result of accelerating numbers
of patents being granted, but suggests a more fundamental change in
the distribution of patents that are being issued. 

\section{Conclusions}%
\label{sec:conclusions}

We have presented a stochastic kinetic model for patent citation
networks.  Though a complex process underlies each decision by a
patent applicant or examiner to cite a particular patent, the average
citation behavior takes a surprisingly simple form.  The citation
probability can be approximated quite well by the ratio of an
``attractiveness function'', $A(k,l)$, which depends on
the in-degree, $k$, and age in patent numbers, $l$, of the cited patent,
and a time-dependent normalization factor, $S(t)$, which is independent
of $k$ and $l$. 

We introduced a method to extract the $A(k,l)$ and $S(t)$ functions of
a growing 
network from a specification of the connection history. We applied
this technique to the patent citation network and, though no
assumptions were made as to the functional form of $A(k,l)$, 
the measured $A(k,l)$ function was well described by two
approximately separable processes: preferential attachment as a
function of in-degree, $k$, and power law age dependence. The interplay
of these two processes, along with the growth in the number of
citations made by each patent, governs the emerging structure of the
network.  Particularly noteworthy are our finding that the
preferential attachment is super-linear, implying that patents are
highly stratified in ``citability'', and our finding of
a power law tail in the age dependence even for small $k$, indicating
not only that some patents remain important for very long times, but
also that even ``dormant'' patents can re-emerge as
important after long delays. 

We also used our technique to investigate the time dependence of the
growth kinetics of the patent citation network.  Overall, we find that
the increasing number of patents issued has been matched by increasing
citations made by each patent, so that the chance that a new patent
will be cited in the next time period has even increased over time.
This result suggests that on average patents are not becoming less
``citable''.  However, we also find that there has been
a change in the underlying growth kinetics since 1991.  Since 1991,
preferential attachment in the patent system has become increasingly
strong, indicating that patents are more and more stratified, with
fewer and fewer of the patents receiving more and more of the
citations.  A few very important, perhaps ``pioneer'',
patents seem to dominate the citations.  This trend may be consistent
with fears of an increasing patent ``thicket'', in which
more and more patents are issued on minor technical advances in any
given area.  These technically dense patents must be cited by patents
that build upon or distinguish them directly, thus requiring that more
citations be made, but few of them will be of sufficient significance
to merit citation by any but the most closely related patents.
Further work will be needed to understand this change in citation
network kinetics. 

This work is only the beginning.  There are many further applications
of network analysis to the patent citation network that are likely to
bear fruit.  It will be possible, for example, to compare the
structural and kinetic behavior of the network for patents in
different technological areas, to investigate the degree of
relatedness between patents in seemingly disparate technologies, and
to explore more detailed structural indicators, such as clustering
coefficients and correlation functions.  Also, it may be possible to
compare the growth of patent systems internationally, perhaps
providing a means to distinguish between the effects of global
technological change and those of nation-specific legal changes.
Finally, it will be interesting to compare the behavior of the patent
citation network with that of other networks (such as the scientific
journal citations discussed in \cite{redner05}) to gain deeper insight
into the behavior of complex networks in general. 

\pnas{\subsection*{Acknowledgments}}
\orig{\begin{acknowledgments}}
This work was funded in part by the National Science Foundation and
the Hungarian Academy of Sciences under grant INT-0332075 and by the
Henry R. Luce Foundation. K. S.'s research is supported
by the DePaul University College of Law. 
\orig{\end{acknowledgments}}

\orig{\bibliographystyle{plainnat}}
\pnas{\bibliographystyle{pnas}}
\bibliography{patentnet}

\end{document}